\documentstyle[12pt,aps,epsf,tighten,epsfig,amsmath,multicol,prbbib]{revtex}

\normalsize

\newcommand{\eq}{\begin{equation}}
\newcommand{\eeq}{\end{equation}}  
\newcommand{\on}{\omega_n}
\newcommand{\intk}{\int \frac{d^2{\mb q}}{4\pi^2}}
\newcommand{\Gammas}{\tilde{\Gamma}}
\newcommand{\mb}{\mathbf}

\newcommand{\noi}{\noindent}

\def\bk{{\bf k}}

\def\bq{{\bf q}}

\def\b0{{\bf 0}}

\def\mbq{{\mb q}}
\def\mbk{{\mb k}}
\def\mbn{{\mb 0}}

\def\Im{{\rm Im}}

\def\up{\uparrow}
\def\down{\downarrow}

\def\om{\omega}
\def\sg{\sigma}

\def\l({\left(}
\def\r){\right)}

\begin{document}

\title{Pair fluctuation induced pseudogap in the normal phase of the
  two-dimensional attractive Hubbard model at weak coupling}
\author{Daniel Rohe and Walter Metzner \\
{\em Institut f\"ur Theoretische Physik C, Technische Hochschule Aachen} \\
{\em Templergraben 55, D-52056 Aachen, Germany}}
\date{\small\today}
\maketitle
\begin{abstract}

\noi One-particle spectral properties in the normal phase of the
two-dimensional attractive Hubbard model are investigated in the weak
coupling regime using the non-selfconsistent T-matrix approximation. 
The corresponding equations are evaluated numerically directly on the 
real frequency axis. 
For temperatures sufficiently close to the superconducting
transition temperature a pseudogap in the one-particle spectral
function is observed, which can be assigned to the increasing 
importance of pair fluctuations.
\noindent
\mbox{PACS: 71.10.Fd, 71.10.-w, 74.20.Mn} \\
\end{abstract}


\section{Introduction} 
\label{sec:intro}

One of the most striking phenomena in high temperature superconductors
\cite{Gin96} is the pseudogap observed in the single-particle spectral 
density and other quantities way above the critical temperature for 
underdoped samples.\cite{Ran98}
A natural explanation of the pseudogap invokes pairing tendencies 
above the pair condensation scale. 
The reduced dimensionality of the cuprate planes and the short 
coherence length of Cooper pairs in the cuprate superconductors 
both favor an enhanced influence of pairing fluctuations.

A popular model allowing the investigation of pairing in two dimensions 
without complications coming from other, especially magnetic,
fluctuations is the two-dimensional attractive Hubbard model.\cite{Mic90}
At low temperatures this model has a superconducting phase with
s-wave pairing.
The existence of a pseudogap for single-particle and spin excitations
in the normal phase of the attractive Hubbard model has been 
established at intermediate to strong coupling by Quantum Monte Carlo 
(QMC) calculations.\cite{Ran92,Sin96,Sin99,Vil98,All99} 
For very strong coupling the pseudogap is a natural
consequence of binding of fermions in almost local pairs, but the
pseudogap has been shown to appear also at more moderate coupling
strengths, where the momentum distribution function still resembles
a Fermi distribution.
Pseudogap behavior in single-particle spectra of the attractive 
Hubbard model has also been obtained within the selfconsistent 
T-matrix approximation (TMA) at very low densities.\cite{Kyu98}
In the low density limit two-particle bound states form already at 
weak coupling in two dimensions,\cite{Sch89} and the chemical 
potential can move into the gap between the energy of a bound state 
and the single-particle band.
At a band-filling of about ten percent and intermediate coupling 
only some weak tendencies towards gap formation appear within the 
self-consistent TMA,\cite{Mic95} and at larger densities standard 
Fermi liquid behavior seems to be restored completely at weak and 
moderate coupling.\cite{Kyu98,Fre92}

In this work we analyze how pairing fluctuations affect the 
spectral function for single-particle excitations in the attractive
Hubbard model at {\em weak coupling}\/ and intermediate (not very
small) densities.
To this end, we compute the self-energy and the spectral function
within the {\em non-selfconsistent}\/ T-matrix approximation.
Using a real-frequency formulation of the T-matrix equations
\cite{Kyu98} and exploiting Kramers-Kronig relations we 
calculate the resulting spectral function with unprecedented
accuracy and energy resolution.
In this way we are able to detect a dramatic drop of spectral
weight at the Fermi level for temperatures close to the 
superconducting transition temperature even at weak coupling.
This pseudogap behavior is caused exclusively by strong pairing 
fluctuations with small center-of-mass momenta precursing the 
superconducting instability.
It can therefore be captured qualitatively within a time-dependent 
Ginzburg-Landau expansion of these fluctuations.\cite{Cap99} 
The gap is not related to two-particle bound states with large
center-of-mass momenta.
Our results are in complete agreement with a classical (i.e.\ 
thermal) pairing fluctuation scenario which has been proposed 
recently as an explanation for pseudogap behavior in the 
two-dimensional Hubbard model at intermediate coupling 
strength.\cite{Vil98}

\section{Method} 
\label{sec:method}

The Hubbard Hamiltonian is given by
\begin{equation} \label{eq:hm_def}
 H \: = \: -\,t\sum_{\langle i,j\rangle,\sg}
 c^{\dagger}_{i\sg} c^{\phantom{\dagger}}_{j\sg} 
 \:+\: U\,\sum_i n_{i\up} \, n_{i\down}
\end{equation}
where $c^{\dagger}_{i\sg}$ and $c^{\phantom{\dagger}}_{i\sg}$ are
the usual creation and annihilation operators for fermions with
spin $\sg$ on site $i$, and
$n_{i\sigma} = c^{\dagger}_{i\sg} c^{\phantom{\dagger}}_{i\sg}$
is the corresponding number operator.
The notation $\langle i,j\rangle$ indicates the restriction of the 
hopping amplitude $t$ to nearest neighbors on the lattice, leading to
the familiar form
$\epsilon^0_{\mb k} \: = \: -2\,t\,(\cos k_x + \cos k_y)$ for the free dispersion relation.
For the {\em attractive}\/ Hubbard model the coupling constant
$U$ is negative.\cite{Mic90}

\begin{figure}
\center
\epsfig{file=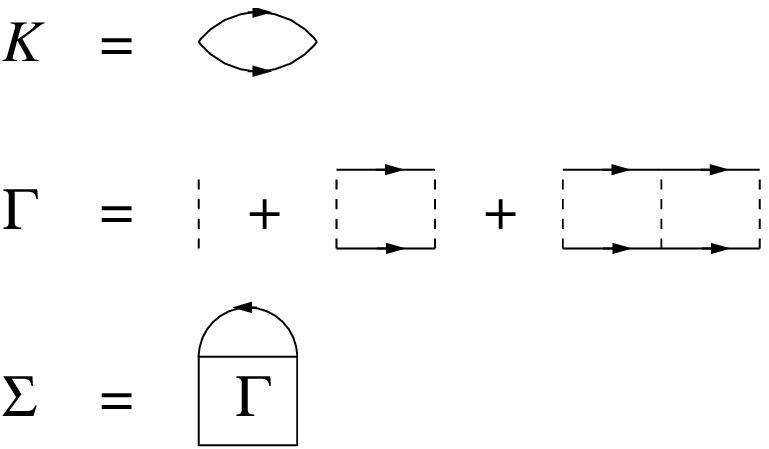,width=8.5cm}
\caption{Pair propagator, vertex function and self--energy within the 
 T-matrix approximation.}
\label{fig:nta}

\end{figure}

In this work physical properties are calculated using the 
non-selfconsistent T-matrix approximation (nTMA) described 
diagrammatically in Fig.\ \ref{fig:nta}. 
Although standard diagrammatic perturbation theory at finite temperatures
is originally formulated for Matsubara frequencies on the imaginary
axis, it is possible to continue the T-matrix equations analytically
to the real frequency axis before solving them numerically directly
for real frequencies.\cite{Kyu98,Sch96} 
Thereby the difficult and ill-posed problem of analytical continuation 
of numerical results from imaginary to real frequencies is avoided. 
The retarded real frequency self-energy within nTMA is then given by
\begin{eqnarray} \nonumber
 \Sigma(\omega + i0^+,{\mb k}) \:=\: \intk \int \frac{d\epsilon}{\pi} \,
 \big\{&-& f(\epsilon)\,\Gamma(\epsilon+\omega +
 i0^+,{\mb q})\,\text{Im}G^0(\epsilon + i0^+,{\mb q} - {\mb k}) \\
 &+& b(\epsilon)\,\text{Im}\Gamma(\epsilon + i0^+,{\mb
 q})\,G^0(\epsilon-\omega
 -i0^+,{\mb q} - {\mb k})
 \big\} 
\label{eq:sigma1}
\end{eqnarray} 
with the vertex function
\begin{equation}
\Gamma(\omega + i0^+,{\mb q}) \:=\: \frac{U}{1 - U K(\omega +
  i0^+,{\mb q})} 
\label{eq:vertex1}
\end{equation}
resulting from the sum over all ladder diagrams with the 
pair-propagator
\begin{equation} 
 K(\omega + i0^+,{\mb q})\:=\: \int \frac{d^2{\mb k}}{4\pi^2}\,
 \frac{f(\mu - \epsilon^0_{{\mb q}-{\mb k}})
 - f(\epsilon^0_{{\mb k}} - \mu)}{\omega - \epsilon^0_{{\mb k}} -
 \epsilon^0_{{\mb q}-{\mb k}} + 2\mu
 + i0^+} 
 \label{eq:bubble1} 
\end{equation}
In the above equations 
$f(\epsilon) = (\exp (\beta\epsilon) + 1)^{-1}$ 
denotes the Fermi function, 
$b(\epsilon) = (\exp (\beta\epsilon) - 1)^{-1}$
the Bose function, 
and $G^0$ is the propagator of the non-interacting system. 

In principle the Fermi surface is (slightly) deformed by interactions.
To avoid the computationally expensive self-consistent determination
of the interacting Fermi surface, we remove Fermi surface shifts by
subtracting ${\rm Re} \Sigma(0,\bk_F)$ from the self-energy 
$\Sigma(\om,\bk)$ entering the expression for the spectral function, 
where $\bk_F$ is a suitable projection of $\bk$ 
onto the (non-interacting) Fermi surface. 
The major effect of this is merely a shift of the chemical potential.

The one-particle spectral function is then obtained as
\begin{equation} 
\label{eq:spectral1}
 A(\omega,\mbk) \:=\: \frac{-2\,\Sigma''(\omega + i0^+,\mbk)} 
 {\left[\omega - (\epsilon^0_{\mbk} - \mu) -
 (\Sigma'(\omega,\mbk)-\Sigma'(0,\mbk_F)) \right]^2 + 
 \left[\Sigma''(\omega + i0^+,\mbk)\right]^2} \quad.
\end{equation}
where $\Sigma'$ and $\Sigma''$ are the real and imaginary parts of
the self-energy, respectively.

We note that eq.\ (\ref{eq:sigma1}) is valid only as long as the  
analytic continuation of the vertex function $\Gamma(z,{\mb q})$ does 
not have poles away from the real frequency axis.
This is discussed in more detail in Appendix \ref{app:nta_real},
where equations (\ref{eq:sigma1})-(\ref{eq:bubble1}) are derived. 
It turns out that poles in the vertex function appear in the complex 
plane if the temperature drops below a critical value $T_c$ at which 
the denominator of the vertex function, i.e.\ $\,1 - U K(0,{\mb 0})$,
vanishes.
This is simply the Thouless-criterion for a superconducting instability.
\cite{Tho60}
The connection between poles of the vertex function in the complex 
plane and the onset of superconductivity has already been pointed out
previously.\cite{Ran89,Fuk91}
For the nTMA the critical temperature is identical to the BCS 
critical temperature $T_c^{BCS}$.\cite{Tho60} 
One must confirm that $T > T_c$ when using eq.\ (\ref{eq:sigma1}) 
to ensure that the system is in the normal phase.

Equations (\ref{eq:sigma1})-(\ref{eq:bubble1}) are evaluated
numerically. 
The computation of the imaginary part of the pair propagator requires
only a one-dimensional numerical integration, since $\Im G^0$ is
proportional to a $\delta$-function. 
The real part is then obtained from the Kramers-Kronig relation. 
The calculation of the imaginary part of the self-energy 
requires a two-dimensional integration, and the real part can again 
be obtained from the Kramers-Kronig integral.
In this way the self-energy and spectral function can be computed 
with very high accuracy and resolution.
The spectral sum rule is satisfied with an accuracy of the order
$10^{-5}$ in our results.
A more detailed description of the numerical treatment of equations
(\ref{eq:sigma1})-(\ref{eq:bubble1}) is given in Appendix 
\ref{app:num_eval}.

For sufficiently large momenta $\mbq$ the vertex function 
$\Gamma(\omega,\mbq)$ has poles on the real frequency axis, which
signal the formation of bound two-particle states.
Their contribution to the self-energy is investigated separately 
in order to distinguish between their influence and all other effects. 
In Appendix \ref{app:tbs} we show that the domain $\text{M}^{\text{TBS}}$ in 
$\mbq$-space for which these bound states exist is given by the 
condition $-2\,t\,\left[ \cos(q_x/2) + \cos(q_y/2) \right] > \mu$,
i.e. the region outside a ``doubled'' Fermi surface.
Note that in the case of an isotropic dispersion relation the condition 
for bound states is simply $|\mbq| > 2\,k_F$.\cite{Sch89}

\section{Results} 
\label{sec:results}

We now present results for the self-energy and the spectral
function. 
The electron density is usually fixed at the intermediate density 
$n = n_{\up} + n_{\down} = 1/2$ (quarter-filling), unless
explicitly stated otherwise.
The hopping amplitude is always chosen to be $t = 1$ for 
simplicity, corresponding to a non-interacting band-width
$w_0 = 8$.

We first establish the existence of a pseudogap in the spectral 
function for a {\em weak}\/ attraction $U = -1.728$, for which the 
corresponding (BCS) transition temperature $T_c = 0.05$ is rather low, 
and discuss its origin.
Since we focus on temperatures close to $T_c$ we use the 
{\em reduced temperature}\/ $\tau_r = (T-T_c)/T_c$\/ to parametrize 
$T$. 

\begin{figure}
\center
\epsfig{figure=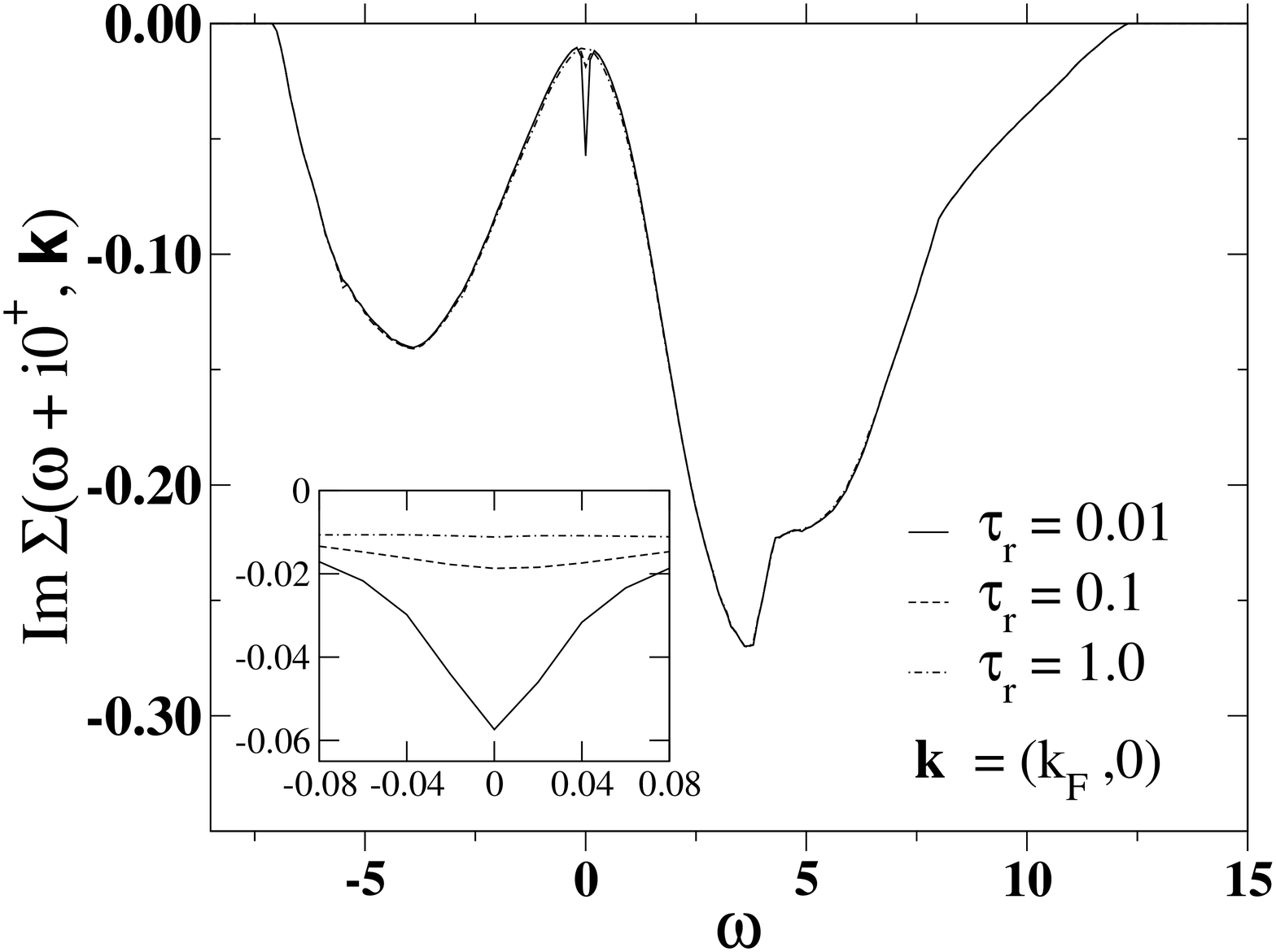,width=.48\linewidth}
\epsfig{figure=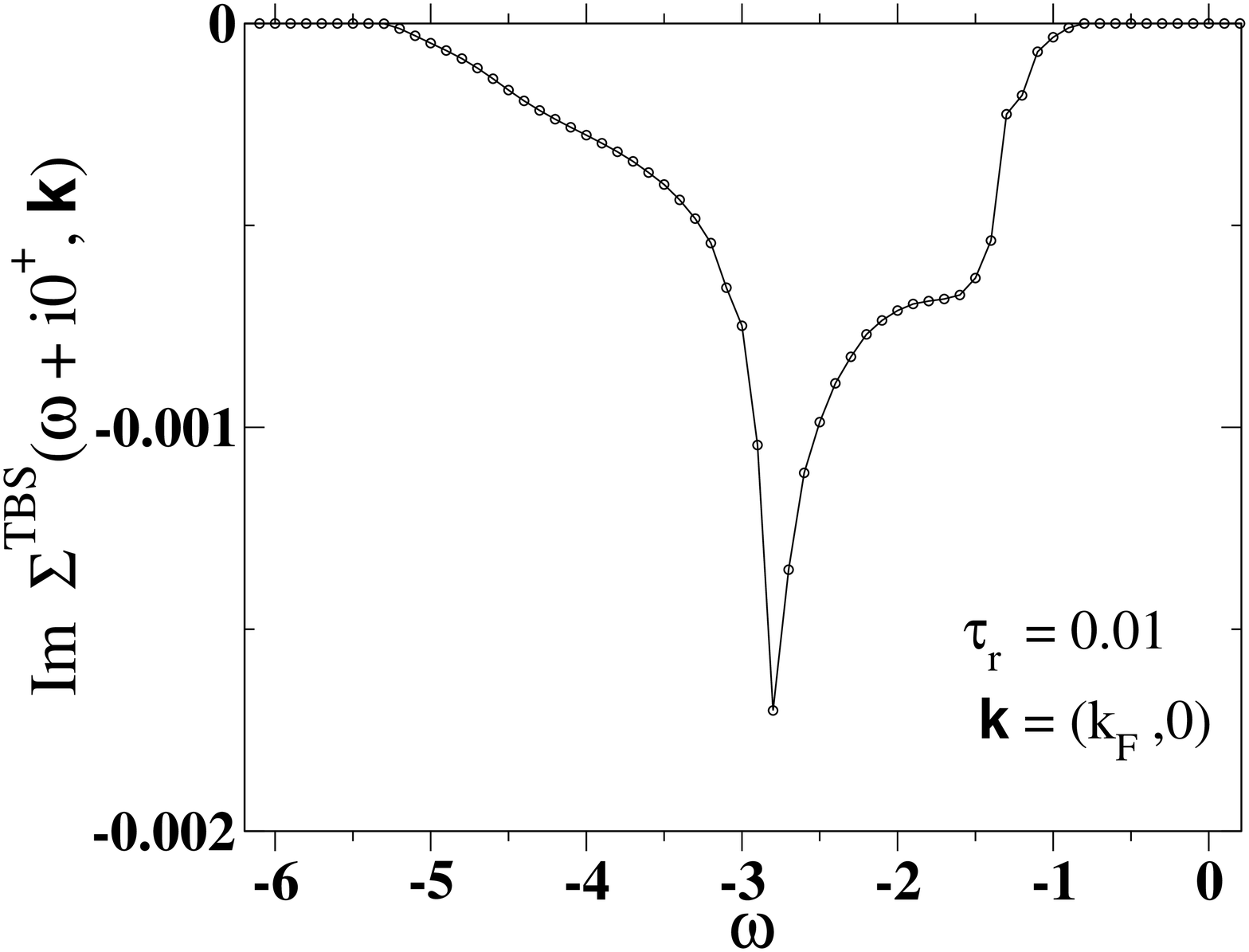,width=.48\linewidth}
\caption{Self-energy at $U = -1.728$
  for $\mbk = (k_F,0)$ and different temperatures (left), 
  and contribution from large $\mbq$ two-particle bound states
  for $\tau_r = 0.01$ (right).}
\label{fig:im_sigma}
\end{figure}

Figure \ref{fig:im_sigma}\/ shows the imaginary part of the
self-energy $\Sigma''(\omega,\mbk)$
for a Fermi wave vector on the $k_x$-axis, $\bk = (k_F,0)\,$, 
at various temperatures, with the contribution $\,\Sigma_{TBS}''\,$ 
from large $\mbq$ two-particle bound states displayed separately for 
the case $\tau_r=0.01$. 
It can clearly be seen that upon decreasing $\tau_r$ a sharp peak
develops in $\,\Sigma''\,$ at $\,\omega = 0\,$. 
This leads to a suppression of spectral weight at 
$\,\omega = 0\,$ (see below) and thus to deviations from Fermi-liquid 
behavior in the normal phase close to $\,T_c\,$.
The origin of this peak is the divergence of the vertex function
$\Gamma(\omega,\bq)$ at $\,\omega = 0,\,\bq = 0 \,$ in the
limit $\,T \to T_c\,$.
Stated in more physical terms, the peak is thus caused by strong 
pair fluctuations with small center-of-mass momenta in the vicinity 
of the superconducting instability.

The bound state contribution $\,\Sigma_{TBS}''\,$ is found to be 
negligible near $\,\omega = 0\,$ and only exhibits a 
rather weak peak at $\,\omega \approx 2\,\mu\,$, which does not 
affect the spectral properties near the Fermi level.
This is is contrast to the scenario proposed by Schmitt-Rink et 
al.,\cite{Sch89} which may be realized only at very low density.
The fact that these bound states do not generically lead to a 
breakdown of Fermi liquid behavior has been pointed out already 
by Fr\'esard et al.\cite{Fre92} and by Serene.\cite{Ser89}

\begin{figure}
\center
\epsfig{figure=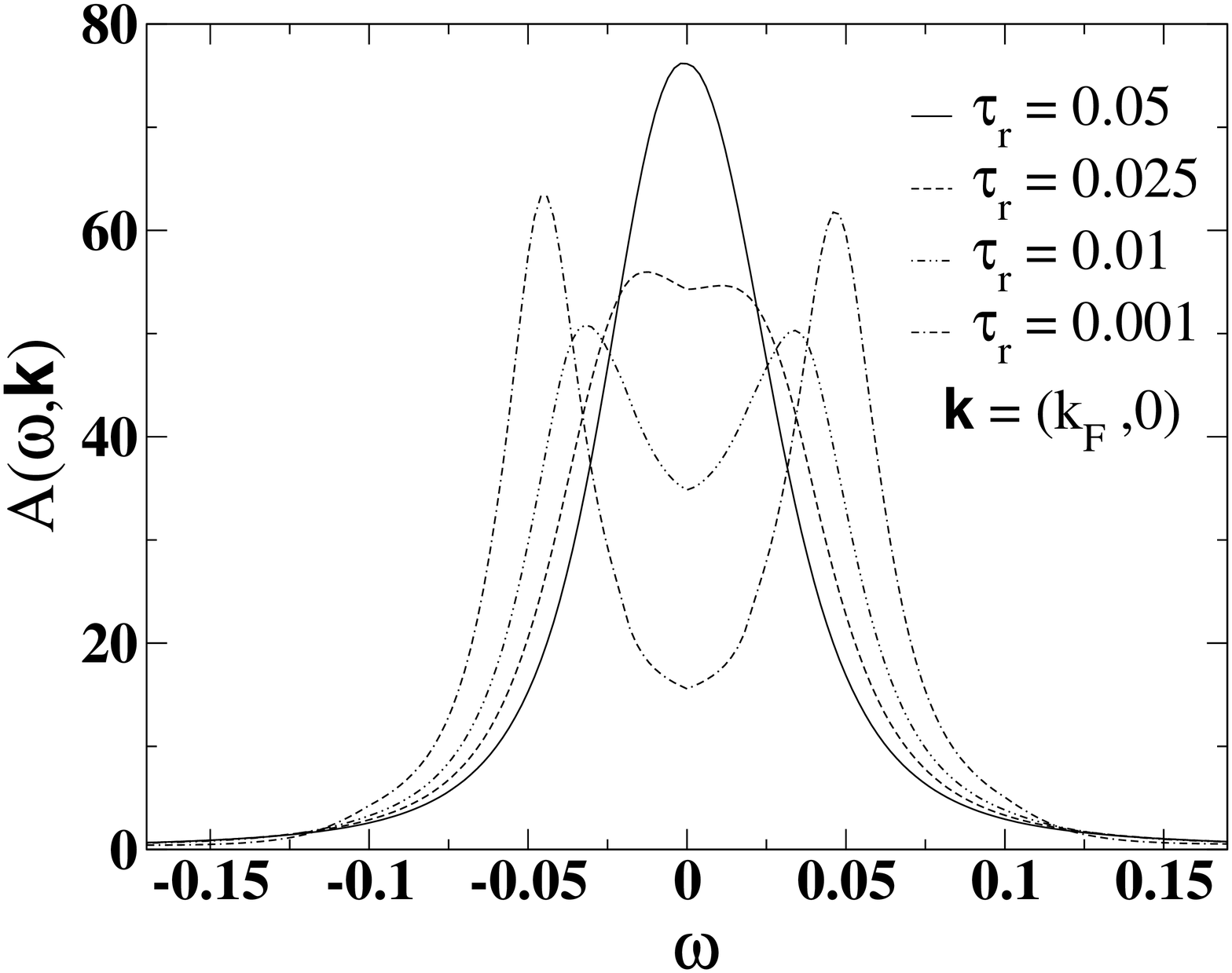,width=8.5cm}
\caption{One-particle spectral function $A(\omega,\mbk)$ at
 $\mbk = (k_F,0)$ 
 for $U = -1.728$ ($T_c = 0.05$)
 and different values of the reduced temperature $\tau_r$.}
\label{fig:spectral1}
\end{figure}

\begin{figure}
\center
\epsfig{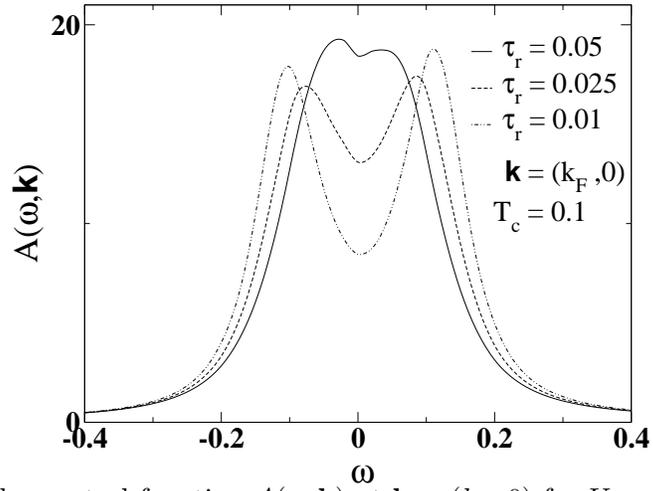}
\caption{One-particle spectral function $A(\omega,\mbk)$ at 
 $\mbk = (k_F,0)$ for $U = -2.034$ ($T_c = 0.1$)
 and different values of the reduced temperature $\tau_r$.}
\label{fig:spectral2}
\end{figure}

\begin{figure}
\center
\epsfig{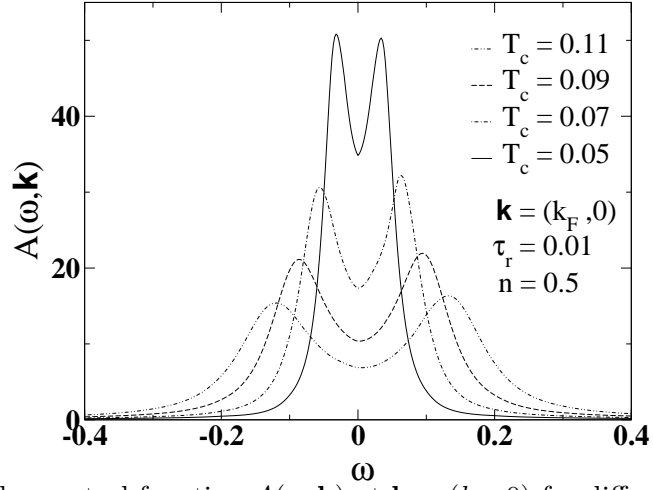}
\caption{One-particle spectral function $A(\omega,\mbk)$ at
 $\mbk = (k_F,0)$ for
 different coupling strengths (i.e.\ various $T_c$)
 and fixed reduced temperature $\tau_r = 0.01$.}
\label{fig:spectral3}
\end{figure}

\begin{figure}
\center
\epsfig{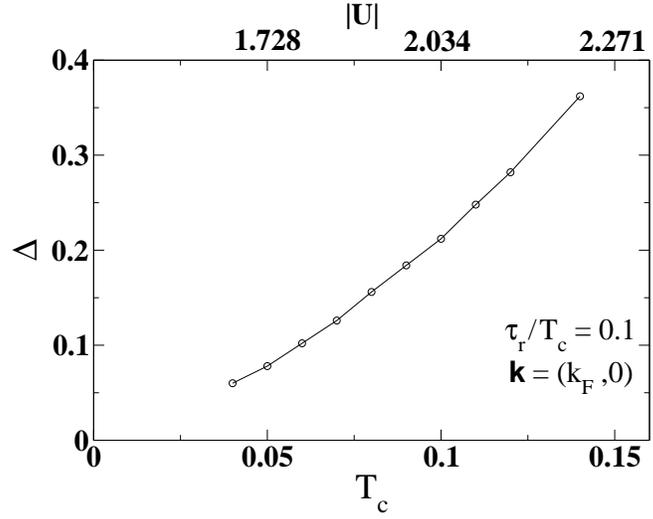}
\caption{Size of the pseudogap as a function of $T_c$ for fixed
 ratio $\tau_r/T_c = 0.1$ at $\mbk = (k_F,0)$.}
\label{fig:gap}
\end{figure}

The resulting spectral function $\,A(\omega,\mbk)\,$ is shown in 
Fig.\ \ref{fig:spectral1} for the wave vector $(k_F,0)\,$.
A pronounced pseudogap forms for sufficiently small reduced
temperatures $\tau_r < 0.01$. 
For stronger attraction both the width of the pseudogap and the
temperature window above $T_c$ where the spectral weight at the
Fermi level gets suppressed increase, as expected.
In Fig.\ \ref{fig:spectral2} we show $\,A(\omega,\mbk)\,$ at 
$\mbk = (k_F,0)$ for a slightly bigger coupling strength 
$U_c = - 2.034$, corresponding to a critical temperature $T_c = 0.1$, 
which is twice as high as before. 
A comparison with Fig.\ \ref{fig:spectral1} shows that the reduced
temperature scale, at which the gap forms, is roughly twice as big
as in the previous case, i.e.\ the absolute temperature window for 
pseudogap behavior is four times larger now.
The width of the pseudogap increases by more than a factor two, 
upon doubling $T_c$.
The systematic evolution of the pseudogap in the spectral function 
with increasing $T_c$ (i.e.\ increasing $|U|$) at fixed $\tau_r = 0.01$ 
is shown in Fig.\ \ref{fig:spectral3}.
The width of the pseudogap increases monotonously with increasing
$|U|$, as expected, and for fixed $\tau_r$ it also becomes more
pronounced.
In Fig.\ \ref{fig:gap} we plot the width of the pseudogap, defined
as the distance between the two maxima in $\,A(\omega,\mbk)\,$,
as a function of $T_c$ (or $|U|$), keeping the {\em ratio}\/ 
$\tau_r/T_c$ fixed at $0.1$, such that the pseudogap is equally well 
developed in all cases. 
The gap width increases obviously faster than linearly with $T_c$.

\begin{figure}
\center
\epsfig{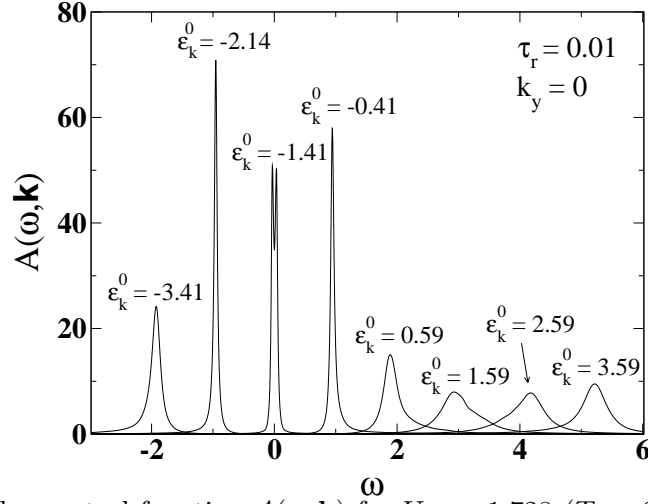}
\caption{One-particle spectral function $A(\omega,\mbk)$ 
 for $U = -1.728$ ($T_c = 0.05$) and $\tau_r = 0.01$ 
 for various wave vectors along the $k_x$-axis.}
\label{fig:spectral4}
\end{figure}

For other wave vectors on the Fermi surface the behavior does not
change much. The pseudogap is however less pronounced for $\mbk$ on 
the diagonal in the Brillouin zone, compared to wave vectors on the
axes (for fixed parameters). 
Note that the van Hove points are still far from the Fermi surface
at quarter-filling. 
For wave vectors away from the Fermi surface the double peak structure
in the spectral function disappears and transforms into a single peak, 
as shown in Fig.\ \ref{fig:spectral4} for the same parameters as in 
Fig.\ \ref{fig:spectral1}.

\begin{figure}
\center
\epsfig{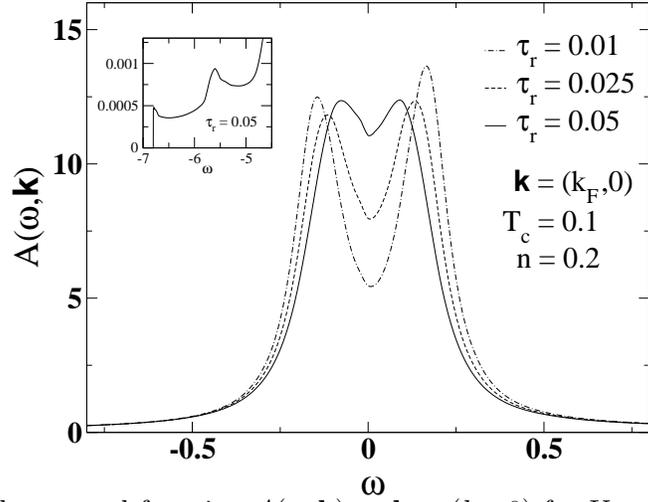}
\caption{One-particle spectral function $A(\omega,\mbk)$ at 
 $\mbk = (k_F,0)$ for $U = -2.667$ ($T_c = 0.1$) 
 and different values of the reduced temperature $\tau_r$;
 the density is $n=0.2$ here.
 The inset shows the tiny peak in the negative frequency tail
 of $A(\omega,\mbk)$ caused by large-$\mbq$ bound states.}
\label{fig:spectral5}
\end{figure}

To see how the spectral function changes at lower densities, we now
switch to a filling factor one tenth ($n = 0.2$).
In Fig.\ \ref{fig:spectral5} we show results for $\,A(\omega,\mbk)\,$ 
at $\mbk = (k_F,0)$ for $U = -2.667$, corresponding to $T_c = 0.1$ 
at that density.
The qualitative behavior of the spectral function is the same as
for quarter filling, except for a minute peak with very little
spectral weight at the lower edge of $\,A(\omega,\mbk)\,$ (see 
inset of Fig.\ \ref{fig:spectral5}). 
This peak is due to the contribution of poles in the vertex function 
at large momenta associated with two-particle bound states. 
Satellite peaks in the spectral function caused by bound states 
have also been obtained earlier within the self-consistent T-matrix
approximation.\cite{Kyu98,Fre92}  
At quarter-filling the bound states contribute to the spectral 
function at an energy which lies within the energy range of the much 
larger contribution from scattering states, and is therefore
hidden by the latter.

A pseudogap in the single-particle excitation spectrum of the 
two-dimensional attractive Hubbard model at finite (not very small) 
density has been found previously only at intermediate and strong 
coupling strengths, namely by Quantum Monte Carlo 
calculations \cite{Ran92,Sin96,Sin99,Vil98,All99} and within a
renormalized T-matrix approximation.\cite{Kyu00}
At weak coupling the temperature window for pseudogap behavior is
very small, and $\bk$ has to be very close to the Fermi surface for
a gap to appear in the spectral function $A(\omega,\bk)$. 
Hence it is very hard to resolve the gap at weak coupling in all 
those numerical techniques for which only systems with finite size 
can be treated in practice.

Thermal pairing fluctuations have already been invoked by Vilk et 
al.\ \cite{Vil98} to account for pseudogap behavior at intermediate 
coupling strengths. 
These authors showed that classical pairing fluctuations can lead
to pseudogaps if the correlation length $\xi$ of these fluctuations 
is not only large, but also larger than the thermal de Broglie wave 
length $v_F/T$, where $v_F$ is the Fermi velocity.
Indeed, we observed at quarter-filling, where the Fermi velocity 
varies already considerably along the Fermi line, that the pseudogap
is more pronounced at Fermi points with a smaller $v_F$.

The dominant long-range part of the pairing fluctuations can be 
described effectively using a Ginzburg-Landau expansion of the 
(diverging) vertex function. 
Capezzali and Beck \cite{Cap99} have recently shown that such an 
expansion captures the singular self-energy contributions which
lead to a pseudogap in the spectral function in two dimensions.

\section{Conclusion}
\label{sec:conclusion}

We have investigated one-particle spectral properties of the 
two-dimensional attractive Hubbard model at weak coupling in the 
normal phase close to the superconducting transition temperature, 
using the non-selfconsistent T-matrix approximation.
For temperatures sufficiently close to the transition temperature the
increasing influence of pair fluctuations induces a pseudogap in the
one-particle spectral function for wave vectors near the Fermi
surface. 
Two-particle bound states with large momenta do not affect the
spectral properties near the Fermi level.
Our results agree qualitatively with those obtained from 
Monte Carlo simulations \cite{Ran92,Sin96,Sin99,Vil98,All99} and
within a renormalized T-matrix approximation \cite{Kyu00} for 
intermediate to strong interaction strengths, and show that a 
pseudogap also forms for a {\em weak}\/ attraction in two dimensions.
The results show that the general pair fluctuation scenario proposed
by Vilk et al.\ \cite{Vil98} works also at weak coupling. They agree  
qualitatively with results obtained recently within a Ginzburg-Landau 
treatment of such fluctuations.\cite{Cap99} 
Hence, the spectral properties of the two-dimensional attractive
Hubbard model seem to evolve smoothly when moving from weak to 
strong coupling, that is there seems to be no critical coupling
strength at which the behavior changes {\em qualitatively}.
The size of the energy window where pairing fluctuations are 
important becomes of course very small at weak coupling, such
that a high numerical accuracy is required to see the effect.
In two dimensions the behavior is thus different than in 
three-dimensional \cite{Jan97} and infinite-dimensional \cite{Kel01} 
systems, where gaps open in the normal state only for a sufficiently 
strong attraction. 

It is clear that the non-selfconsistent T-matrix approximation
breaks down in the limit $T \to T_c$, since the propagator is
dramatically renormalized at the Fermi level. 
Contributions not contained in the TMA may modify the shape of 
the pseudogap in the limit $T \to T_c$, even at weak coupling.
The selfconsistent TMA will hardly provide a better approximation
since vertex corrections are likely to become important as soon as 
self-energy insertions contribute significantly.\cite{Kyu00}
None of the TMA versions captures the Kosterlitz-Thouless physics
of the true superconducting phase transition.

Improved theories should take vertex and self-energy corrections
into account on an equal footing. 
An improvement of the TMA which agrees better with QMC results 
for the attractive Hubbard model at intermediate coupling strengths 
has been proposed very recently by Allen and Tremblay \cite{All00}
and evaluated in detail by Kyung et al.\cite{Kyu00}
The bare interaction is replaced by a renormalized interaction
which is determined by consistency requirements in this approach.
An advantage of this renormalized scheme compared to the standard
T-matrix is that the temperature window for normal state pseudogap 
behavior becomes larger, in agreement with QMC results at 
intermediate coupling. 
However, $T_c$ is suppressed too much by fluctuations in this 
approach, namely down to zero in two dimensions, and the expected 
Kosterlitz-Thouless behavior (at densities $n \neq 1$) is not 
captured.
The true $T_c$ indeed vanishes only in the half-filled limit 
($n \to 1$), where superconductivity would have to break a larger 
$SO(3)$ symmetry, which leads to stronger order parameter 
fluctuations and thus to a further enhancement of the pseudogap 
regime.\cite{All99}
By contrast, in the standard T-matrix approximation (self-consistent 
or not) the order parameter fluctuations are always underestimated 
since the phase transition is mean-field like, and the pseudogap 
regime appears thus too small. 

Functional renormalization group methods, where vertex corrections 
are automatically included, should provide another promising
approach to pseudogap physics. 
Such methods have been applied recently to the repulsive Hubbard 
model,\cite{RG} but not yet to the attractive model.

\vskip 1cm

\noindent
{\bf Acknowledgements:} \\
We are grateful to B. Kyung and A.-M.S. Tremblay for very valuable
correspondence. 
This work has been supported by the Deutsche Forschungsgemeinschaft
within the Sonderforschungsbereich 341.

\begin{appendix}


\section{Derivation of T-Matrix equations on the real axis} 
\label{app:nta_real}

How to continue approximation schemes resulting from a resummation
of Feynman diagrams to the real frequency axis has been pointed out 
for example in Ref.\ \onlinecite{Sch96}, where the case of the fluctuation 
exchange approximation is treated explicitly.
The selfconsistent T-matrix equations for real frequencies
have been presented (without derivation) already in Ref.\ 
\onlinecite{Kyu98}. Our equations could be obtained from the latter
replacing simply $G$ by $G^0$ on the right hand side.
Nevertheless we will now discuss the analytic continuation of
the nTMA in some detail, in order to make the paper self-contained, but 
in particular to clarify under which conditions this
continuation can be done.

\begin{figure}
\center
\epsfig{figure=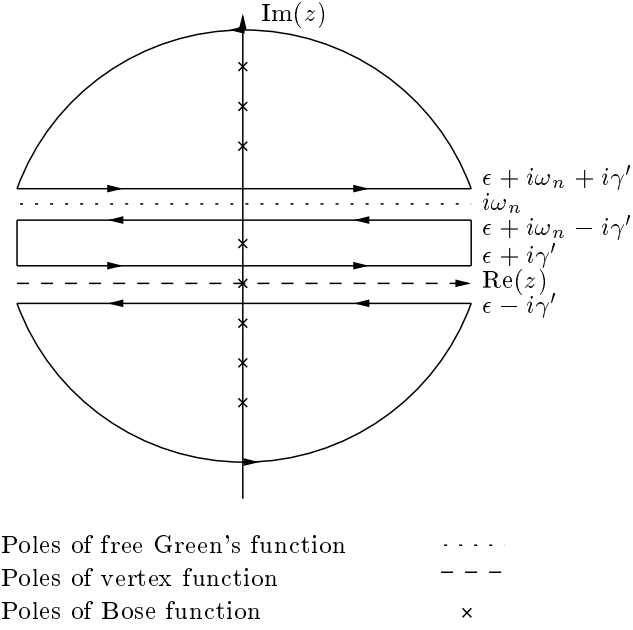,width=8.5cm}
\caption{Integration contour used to derive equation 
 (\ref{eq:sigma_contour}).}
\label{fig:contourselbst}
\end{figure}

The non-selfconsistent T-matrix equations for the attractive
Hubbard model with Matsubara frequencies on the imaginary axis are 
given by 
\begin{eqnarray}
 \Sigma(i\on,\mbk)&=& T \sum_{\nu_n} e^{i\nu_n 0^+}
 \int \frac{d^2\mbq}{4\pi^2}\,
 \Gamma(i\nu_n,\mbq) \, G^0(i\nu_n - i\on,\mbq-\mbk) 
 \label{eq:sigma_mats} \\[2mm]
 \Gamma(i\nu_n,\mbq) &=& \frac{U}{1 - U K(i\nu_n,\mbq)} 
 \label{eq:vertex_mats}\\[2mm]
 K(i\nu_n,\mbq) &=& -T \sum_{\on} \int \frac{d^2\mbk}{4\pi^2}\, 
 G^0(i\on,\mbk) \, G^0(i\nu_n-i\on,\mbq-\mbk) 
 \label{eq:bubble_mats}
\end{eqnarray}
After carrying out the Matsubara sum in the pair propagator 
(\ref{eq:bubble_mats}), the analytic continuation to the real
axis (with the correct large frequency behavior) is obtained by 
simply substituting $i\nu_n \mapsto \nu + i\delta$, which 
yields
\begin{equation}\label{eq:bubble_real}
 K(\nu + i\delta,\mbq) = \int \frac{d^2\mbk}{4\pi^2}\, 
 \frac{f(\mu - \epsilon^0_{\mbq-\mbk}) - f(\epsilon^0_{\mbk} - \mu)}
 {\nu - \epsilon^0_{\mbk} - \epsilon^0_{\mbq-\mbk} + 2\mu + i\delta}
\end{equation} 
The continuation of the vertex function follows immediately as 
\begin{equation}\label{eq:vertex_real}
 \tilde{\Gamma}(\nu + i\delta,{\mb q}) \: := 
 \: \Gamma(\nu + i\delta,{\mb q}) - U \: =
 \: \frac{U^2\,K(\nu + i\delta,{\mb q})}
         {1 - U K(\nu + i\delta,{\mb q})}
\end{equation}
The term $U$ has been subtracted from $\Gamma$ in order to isolate 
the Hartree term in the self-energy

\begin{equation} \label{eq:sigma_mats_2}
 \Sigma(i\on,\mbk) = \Sigma^H + \tilde\Sigma(i\on,\mbk) =
 T \sum_{\nu_n} e^{i\nu_n 0^+} \int \frac{d^2\mbq}{4\pi^2} 
 \left( U + \tilde{\Gamma}(i\nu_n,\mbq) \right) 
 G^0(i\nu_n - i\on,\mbq-\mbk)
\end{equation}
where the (constant) Hartree contribution can be easily summed to 
$\Sigma^H = U n/2$.
The second term in equation (\ref{eq:sigma_mats_2}) is treated using 
the contour integration technique with the integration contour 
shown in Fig.\ \ref{fig:contourselbst}. This yields

\begin{eqnarray} \nonumber
\hspace{-0.5cm} 
 \tilde\Sigma(i\on,\mbk) \,&=&\,
 \Big[\, T \, \tilde{\Gamma}(0,\mbq)\, G^0(-i\on,\mbq-\mbk) 
 \\ \nonumber
 + \: \frac{1}{2\pi i} \int d\epsilon &\big\{&\hspace{0mm}  
 b(\epsilon + i\gamma' + i\on)\,
 \Gammas(\epsilon+i\gamma'+i\on,\mbq)\,G^0(\epsilon+i\gamma',\mbq-\mbk) 
 \\ \nonumber
 &+&\hspace{0mm} b(\epsilon + i\gamma')\,
 \Gammas(\epsilon+i\gamma',\mbq)\,G^0(\epsilon+i\gamma'-i\on,\mbq-\mbk)
 \\ 
 &-& (\gamma' \mapsto -\gamma')
 \big\} \Big] \label{eq:sigma_contour}
\end{eqnarray}
with $\gamma'>0$ being a small positive number eventually taken to 
be zero. 
The term $T\,\tilde{\Gamma}(0,\mbq)\, G^0(-i\on,\mbq-\mbk)$ is not 
covered by any of the contours and is therefore included explicitly.

\begin{figure}
\center
\epsfig{figure=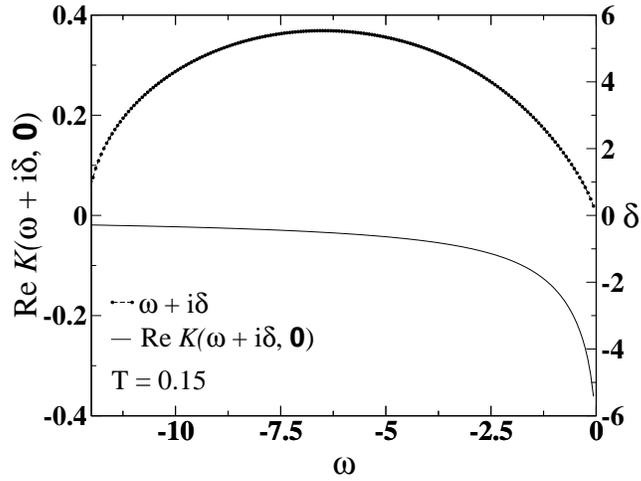,width=8.5cm}
\caption{Zeros of the imaginary part of the pair propagator at $z =
  \omega + i\delta$ in the
  complex plane (upper curve) and real part of the pair propagator at
  these points (lower curve) for $\mbq = 0$ and $T = 0.15$.}
\label{fig:null}
\end{figure}

At this point it is important to notice that for equation 
(\ref{eq:sigma_contour}) to be valid the vertex function 
{\em must not have poles away from the real axis}. 
Such poles appear for $|U| > |U_c(T)|$ (or equivalently for
$T < T_c(U)$), where $U_c$ is determined by the Thouless criterion 
$\,1 - U_c\,K(0,\mbn)=0$. 
To illustrate this, Fig.\ \ref{fig:null} shows the zeros of the 
imaginary part of the pair propagator at $z = \omega + i\delta$ for 
$\mbq = \mbn$ together with the real part at these very points. 
The absolute value of the real part has a maximum at $\omega = 0$. 
Hence, upon increasing the coupling strength a pole first appears
for $|U| = |U_c|$ at $\omega = 0$, and then moves into the complex 
plane for larger $|U|$, along the line defined by the zeros of the 
imaginary part.

Provided that $|U| < |U_c|$ the steps to finalize the analytic
continuation of the self-energy can be summarized as follows:
(i) use the relation $b(z+i\on) =-f(z)$ to replace Fermi
functions by Bose functions,
(ii) make the replacement  $i\on \to \omega + i\gamma$ with $\gamma >
\gamma'$ being a small parameter,
(iii) take the limit $\gamma' \to 0^+$ with $\gamma$ remaining finite
and use $\lim_{\gamma'\to 0^+}\text{Im} \, b(\epsilon + i\gamma') =
-i\pi\,T\delta(\epsilon)$,
(iv) take the limit $\gamma \to 0^+$.
The result of this procedure is 
\begin{eqnarray} \nonumber
\tilde\Sigma(\omega + i0^+,{\mb k}) \:=\:
 \frac{1}{\pi} \intk \int d\epsilon \, \big\{&-& f(\epsilon)\,
 \Gammas(\epsilon+\omega +
 i0^+,{\mb q})\,\text{Im}G^0(\epsilon + i0^+,{\mb q}-{\mb k}) \\
 &+& b(\epsilon)\,\text{Im}\Gammas(\epsilon + i0^+,{\mb q})\,
 G^0(\epsilon-\omega-i0^+,{\mb q}-{\mb k})
\big\} 
\label{eq:selfenergy_app1}
\end{eqnarray} 
Inserting $\,\Gammas = \Gamma - U\,$ and adding $\Sigma^H$ yields 
equation (\ref{eq:sigma1}).


\section{Numerical evaluation of T-Matrix equations}
\label{app:num_eval}

In order to numerically evaluate the pair propagator the following
change of variables is made:
$\xi := \epsilon^0_{\mbk} + \epsilon^0_{\mbq-\mbk} = 
 -a \cos\tilde{k}_x -b \cos\tilde{k}_y$ and $\lambda := \tilde{k}_x$,
where $a = 4t\cos(q_x/2)$, $b = 4t\cos(q_y/2)$, and
$\tilde{k}_i = k_i - q_i/2$.
Integration in k-space is then written as
\begin{equation*}
 \int  d^2\mbk \: \dots \: = \:
 \int \limits^{\xi_2}_{\xi_1} d\xi \,
 \int \limits_{\lambda_1}^{\lambda_2} d\lambda 
 \frac{\sum_{Qu.1}^{Qu.4}}{b \sin \left[
  \tilde{k}_y \left(\xi,\lambda \right) \right]}
 \: \dots
\end{equation*}
where $\xi_1 = - (a + b)$, $\xi_2 \: = \: a + b$, and
\begin{eqnarray*}
 \lambda_1 &=& 0 ,\quad \lambda_2 = 
 \arccos \left( - \frac{\xi + b}{a} \right),
 \qquad  \text{for} \: \: \, - a - b \leq \xi < a - b \\[2mm]
 \lambda_1 &=& 0 ,\quad \lambda_2 = \pi,
 \quad \qquad \qquad \qquad \qquad \text{for} 
 \qquad a - b \leq \xi \leq b - a \\[2mm]
 \lambda_1 &=&  \arccos \left( \frac{b - \xi}{a} \right),
 \quad \lambda_2 = \pi,
 \qquad \: \: \: \text{for} \qquad b - a < \xi \le a - b \quad.
\end{eqnarray*}
with $\sum_{Qu.1}^{Qu.4}$ indicating the summation over all four 
quadrants of the Brillouin zone.
While this does not reduce the numerical effort for finite
$\delta$ in equation (\ref{eq:bubble_real}) it is a great advantage 
in the limit $\delta \to 0^+$ taken in (\ref{eq:bubble1}). 
Since the imaginary part of the integrand is proportional to 
$\delta(\omega + 2\mu - \xi)$, the $\xi$-integral is trivial and  
only a one-dimensional integration needs to be performed numerically to
compute the imaginary part of the pair propagator:  
\begin{equation} \label{eq:bubble_imag}
\text{Im} K(\omega + i\, 0^+,\mbq)  \:=\: -\pi\,  
\int \limits_{\lambda_1}^{\lambda_2} \frac{d\lambda}{4\pi^2} \,
\left. \frac{\sum_{Qu.1}^{Qu.4} \left(f(\mu - \epsilon^0_{\mbq-\mbk})
- f(\epsilon^0_{\mbk} - \mu)\right) }{b \sin \left[
    \tilde{k}_y \left(\xi,\lambda \right) \right]} \: \right|_{\xi =
\omega + 2\mu}
\end{equation}
The real part is then obtained using the Kramers-Kronig relation

\begin{equation} \label{eq:real_real}
\text{Re} K(\omega + i0^+,\mbq)  \:=\:
P \int \frac{d\omega'}{\pi} \, \frac{\text{Im}
 K(\omega'+i0^+,\mbq)}{\omega' -\omega} 
\end{equation}

The calculation of the self-energy follows a similar procedure using
the slightly different transformation
$\xi \: = \: -2\,t (\cos q_x + \cos q_y) \: = \: \epsilon^0_{\mbq}$,
$\lambda \: = \: 2\,t (\cos q_x - \cos q_y)$, 
with the Jacobian
\begin{equation*}
J(\xi,\lambda) = 2 \, \left( \sqrt{256\,t^4 + \xi^4 + \lambda^4 - 2\,
    \left( 16 \, t^2 \lambda^2 + 16 \, t^2 \xi^2 + \lambda^2 \xi^2
    \right)} \right)^{-1} \label{eq:sigma_trafo}
\end{equation*}
\noi after which integration in q-space becomes
\begin{equation}
 \int \frac{d^2\mbq}{4\pi^2} \: \dots \: = \: 
 \int \limits^{4t}_{-4t} \frac{d\xi}{4\pi^2} \,
 \int \limits_{\lambda_1}^{\lambda_2} d\lambda \, J(\xi,\lambda)
 \sum_{Qu.1}^{Qu.4} \: \dots
\end{equation}
with $\lambda_1 = -(4t-|\xi|)$ and $\lambda_2 = (4t-|\xi|)$.
The imaginary part of the self-energy is then given by a 
two-dimensional integral
\begin{eqnarray}
 && \hspace{-1cm} \Im\Sigma(\omega + i0^+,\mbk) \: = 
 \\[3mm] \nonumber 
 &&\hspace{-1cm} \int \limits^{4t}_{-4t} 
 \frac{d\xi}{4\pi^2} \, 
 \left\{f(\xi - \mu) + b(\omega + \xi - \mu) \right\}
 \int \limits_{\lambda_1}^{\lambda_2} d\lambda \, J(\xi,\lambda)
 \sum_{Qu.1}^{Qu.4} \text{Im}\Gammas(\omega + \xi - \mu +
 i0^+,\mbq(\xi,\lambda) + \mbk)
\label{sigma_num_2}
\end{eqnarray} 
and the real part is again obtained using the Kramers-Kronig
relation:
\begin{equation}
\text{Re} \Sigma(\omega + i0^+,\mbk)  \:=\: U\frac{n}{2} \:+\:
 P \int \frac{d\omega'}{\pi}\, \frac{\text{Im}
  \Sigma(\omega' + i0^+,\mbk)}{\omega' -\omega}
\end{equation}
%


\section{Two-particle bound states}
\label{app:tbs}

\begin{figure}
\center
\epsfig{figure=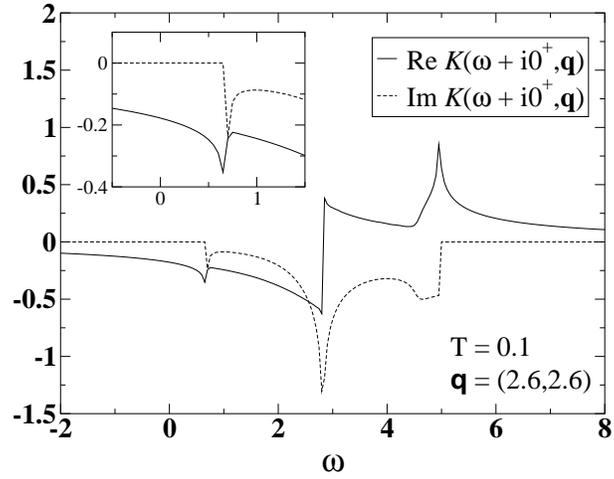,width=8.5cm}
\caption{Frequency dependence of the pair propagator for 
 $\mbq = (2.6,2.6)$ at $T = 0.1$.}
\label{fig:bubble}
\end{figure}

\begin{figure}
\center
\epsfig{figure=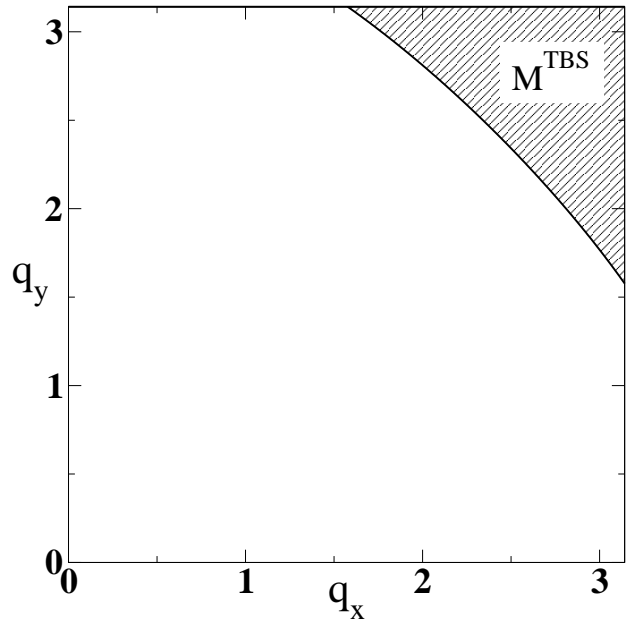,width=8.5cm}
\caption{Domain $\text{M}^{\text{TBS}}$ in which two-particle bound states exist
 (for quarter-filling).}
\label{fig:defTBS}
\end{figure}

To identify the values $\mbq$ for which two-particle bound states exist
we first note that the domain, in which the imaginary part of the
pair propagator (see Fig.\ \ref{fig:bubble}) is non-zero in the limit 
$\delta \to 0^+$, is bounded from below by $\omega_u(\mbq) = 
 -4t[\cos(q_x/2)+\cos(q_y/2)] - 2\,\mu$. 
In the case $\omega_u(\mbq) > 0$ the imaginary part jumps
at $\omega_u(\mbq)$ from $0$ to a finite {\em negative}\/ value. 
This leads to a negative logarithmic singularity in the real part for 
$\omega \to \omega_u(\mbq)^-$ and in turn to the appearance of a
delta-peak in the imaginary part of the vertex function at some 
point $\nu(\mbq)$ with $0 < \nu(\mbq) < \omega_u(\mbq)$, 
representing a two-particle bound state.
The border of the region $\text{M}^{\text{TBS}}$ in which two-particle bound states
exist is defined by $\omega_u(\mbq) = 0$ and is shown for 
quarter-filling in Fig.\ \ref{fig:defTBS}. 
The condition $\omega_u(\mbq) = 0$ reads explicitly
\begin{equation} \label{eq:tbs}
-2\,t\,\left[\cos(q_x/2) + \cos(q_y/2)\right] \:=\: \mu
\end{equation}  
defining a one-dimensional ``surface'' that is identical to a Fermi 
surface ``up-scaled'' by a factor two. 

\end{appendix}


\vfill\eject


\centerline{\Large FIGURES}
\vskip 1cm


\end{document}